\documentclass[fleqn,10pt]{wlscirep}
\usepackage[utf8]{inputenc}
\usepackage[T1]{fontenc}
\usepackage{amssymb,amsmath}
\usepackage{subfig}
\usepackage{graphicx}
\usepackage{color}
\usepackage{hyperref}
\usepackage{listings}
\usepackage{graphicx}
\usepackage{amsthm}
\usepackage{amsfonts}
\usepackage{braket}
\usepackage{qtree}
\usepackage{dsfont}
\usepackage{mathtools}
\usepackage{lipsum, babel}
\usepackage{bbm}
\usepackage{mathtools}
\usepackage{physics}
\usepackage{placeins}
\usepackage{bbold}

\title{The Effect of Classical Optimizers and Ansatz Depth on QAOA Performance in Noisy Devices}

\author[1,2]{Aidan Pellow-Jarman}
\author[2,3]{Shane McFarthing}
\author[1,4,*]{Ilya Sinayskiy}
\author[5,6]{Daniel K. Park}
\author[1,7]{Anban Pillay}
\author[1,2,4,8]{Francesco Petruccione}

\affil[1]{University of KwaZulu-Natal, Durban, South Africa, 3629}
\affil[2]{Qunova Computing Inc, Daejeon, 34051, South Korea}
\affil[3]{Stellenbosch University, Department of Physics, Stellenbosch, 7600, South Africa}
\affil[4]{National Institute for Theoretical and Computational Sciences (NITheCS), South Africa}
\affil[5]{Department of Applied Statistics, Yonsei University, Seoul, 03722, South Korea}
\affil[6]{Department of Statistics and Data Science, Yonsei University, Seoul, 03722, South Korea}
\affil[7]{Centre for Artificial Intelligence Research (CAIR), South Africa}
\affil[8]{Stellenbosch University, School of Data Science and Computational Thinking, Stellenbosch, 7600, South Africa}

\affil[*]{sinayskiy@ukzn.ac.za}

\begin{abstract}
The Quantum Approximate Optimization Algorithm (QAOA) is a variational quantum algorithm for Near-term Intermediate-Scale Quantum computers (NISQ) providing approximate solutions for combinatorial optimiz\-ation problems. The QAOA utilizes a quantum-classical loop, consisting of a quantum ansatz and a classical optimizer, to minimize some cost function, computed on the quantum device. This paper presents an investigation into the impact of realistic noise on the classical optimizer and the determination of optimal circuit depth for the Quantum Approximate Optimization Algorithm (QAOA) in the presence of noise. We find that, while there is no significant difference in the performance of classical optimizers in a state vector simulation, the Adam and AMSGrad optimizers perform best in the presence of shot noise. Under the conditions of real noise, the SPSA optimizer, along with ADAM and AMSGrad, emerge as the top performers. The study also reveals that the quality of solutions to some 5 qubit minimum vertex cover problems increases for up to around six layers in the QAOA  circuit, after which it begins to decline. This analysis shows that increasing the number of layers in the QAOA in an attempt to increase accuracy may not work well in a noisy device.
\end{abstract}
\begin{document}

\flushbottom
\maketitle
%
%
\thispagestyle{empty}


\section*{Introduction}

Many promising quantum algorithms, offering polynomial and exponential speed-ups over their classical counterparts, have been proposed \cite{nielsen}. These algorithm\-s, including Grover's unstructured search algorithm \cite{grover} and Shor's algorithm for finding the prime factors of an integer \cite{shor}, have generated much excitement over the prospects of quantum computing. However, their practical realization is generally accepted to be a long term project due to constraints such as noise and state fidelity \cite{nielsen}. Error-correction schemes that would yield fault-tolerant quantum computers have been devised, but they require quantum compute\-rs with many more qubits than we have available at present \cite{preskill}. 

In the meanwhile, there is significant interest in quantum algorithms that are applicable to noisy intermediate-scale quantum (NISQ) computers currently available  in the near to intermediate future \cite{preskill}. These algorithms are predominantly variational and use hybrid quantum-classical routines to leverage existing quantum resources. They include the Variational Quan\-tum Eigensolvers (VQE) \cite{peruzzo} and the Quantum Approximate Optimization Algo\-rithm (QAOA) \cite{qaoa}. As the number of the qubits in these near-term quantum computers increases, they become increasingly difficult to simulate with classical computers \cite{nielsen}.

The VQE and QAOA algorithms utilize parameterized quantum circuits $U\left(\theta\right)$ to evolve the state of the Hamiltonian $H$ representing the problem of interest. Using the expectation value $\bra{U\left(\theta\right)}H\ket{U\left(\theta\right)}$, a classical optimizer is used to train the parameters of the quantum circuit. In this way, the algorithm uses the ansatz to prepare trial solutions to the problem, and the classical optimizer searches for better approximations of the ideal solutions \cite{peruzzo, qaoa}.

In QAOA, the ansatz is constructed from $p$ layers of exponentiated cost and mixer Hamiltonians obtained from the problem cost definition \cite{qaoa}. As $p\rightarrow\infty$, the solution prepared by QAOA approaches the ideal solution, but, in the context of NISQ computing it is not feasible to utilize such deep circuits due to the effects of noise and state decoherence \cite{qaoa, preskill}. The noise in the NISQ computers also adversely affects the efficacy of the classical optimization procedure. The performance of the classical optimizer has recently been studied on the QAOA \cite{qaoa_opt}, as well as in other variational quantum algorithms \cite{pellow-jarman}. Optimization protocols of several variations of the QAOA have also been recently studied \cite{qaoa_var}.

 In this work, we investigate classical optimizers and circuit depths $p$ to find the optimal optimizer choice and ansatz depth for the minimum vertex cover problem under realistic device noise. We utilized a noise model sampled from the IBM Belem quantum computer to simulate the effects of noise on the efficacy of the algorithm. To the best of our knowledge, this is the first investigation of optimal circuit depth for QAOA with noise.

The remainder of the paper is structured as follows: Section \ref{qaoa_min_vertex} revises the details of the minimum vertex cover problem and the QAOA algorithm, Section \ref{opt} describes the optimizers used, and the test methodology followed in this work, and Section \ref{results} presents our findings and discusses their significance.

\section{Quantum Approximate Optimization Algorithm and Minimum Vertex Cover Problem}
\label{qaoa_min_vertex}
\subsection{Minimum Vertex Cover Problem}
The Minimum Vertex Cover problem is an example of a binary optimization problem that is NP-complete.
A vertex cover of a graph $G = (V, E)$, is a set of vertices $V' \subseteq V$, such that for every edge $e = (u, v) \in E, u \in V' \cup v \in V'$. The minimum vertex cover is a set of vertices $V*$, that is the smallest possible set satisfying the above condition for a given graph $G$. The minimum vertex cover problem is to find a set $V*$.

\begin{figure}
    \includegraphics[width=\textwidth]{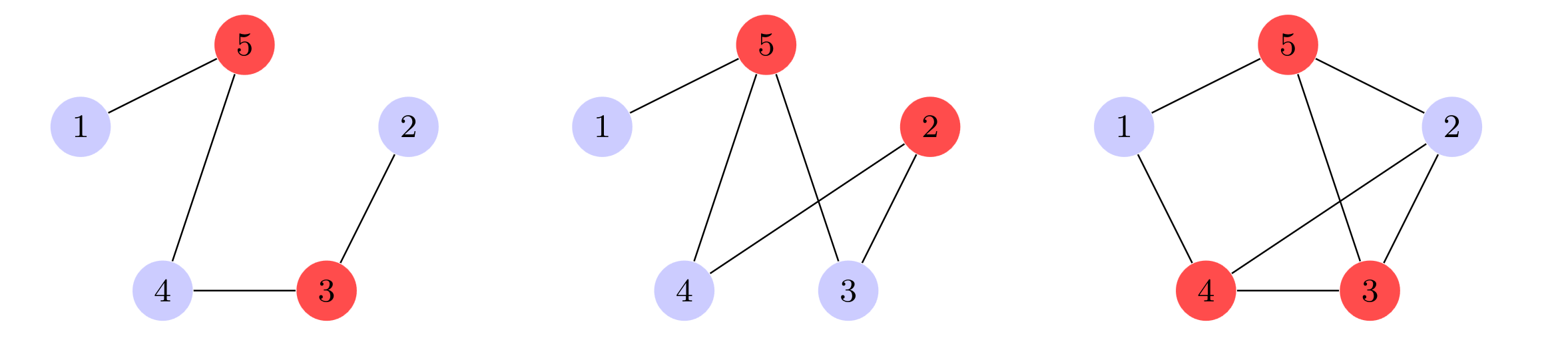}
    \caption{In the three graphs above, the red nodes show the set of vertices forming each graph's respective minimum vertex cover. Each edge in  the graph under consideration must have, at least one vertex in the cover. The cover forms the minimum cover of a graph, when it contains the fewest number of vertices, whilst ensuring each edge is still incident to at least one.}
    \label{graphs_example}
\end{figure}

The minimum vertex cover problem can be formulated as the following binary optimization problem:
\begin{align}
\text{Minimize:} \quad & \sum_{i \in V} x_i \\
\text{Subject to:} \quad & x_i + x_j \geq 1, \quad \forall (i, j) \in E \\
\text{and:} \quad & x_i \in \{0,1\}, \quad \forall i \in V
\end{align}
In Figure \ref{graphs_example}, we provide examples of graphs that illustrate the minimum vertex cover problem.

Binary optimization problems, like the minimum vertex cover problem, have their solutions encoded in a bit string and require an algorithm capable of finding the appropriate bit string to minimize the cost function. For the minimum vertex cover, each bit in the bit string corresponds to a vertex in the problem graph. A bit value of 1 indicates that the vertex is in the cover set, and a bit value of 0 indicates that the vertex is not in the cover set. The QAOA is one such quantum algorithm capable of finding an approximate solution in the form of a bit string, read out of the quantum device directly through measurement. Each qubit corresponds to a vertex in the graph, and the measured value of 0 or 1, forms a bit string solution for the problem.

\subsection{Quantum Approximate Optimization Algorithm}

\begin{figure}
    \includegraphics[width=\textwidth]{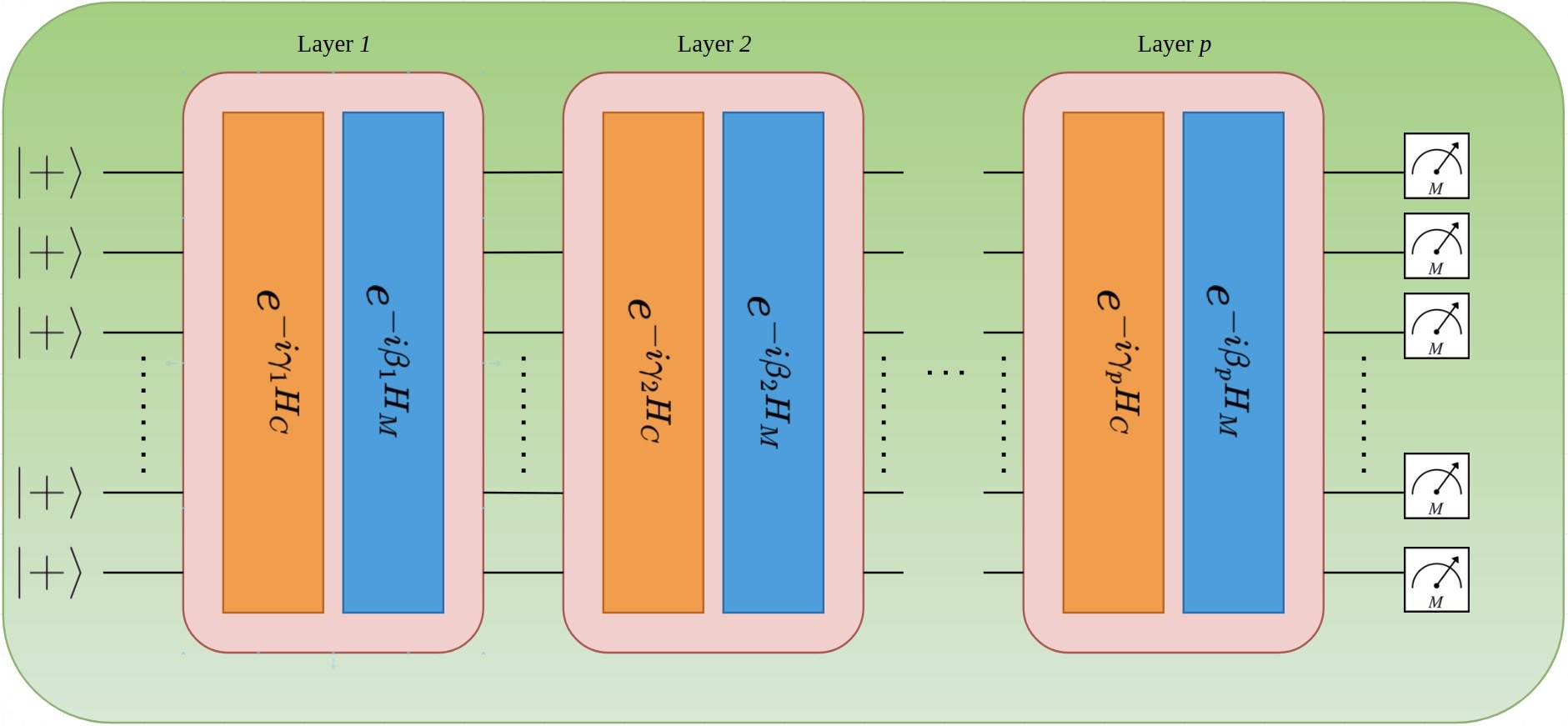}
    \caption{The QAOA circuit consists of \textit{p} layers of the cost and mixer Hamiltonians, $H_C$ and $H_M$ respectively. The initial $\ket{+}^{\otimes N}$ state is prepared and every qubit is measured after applying the QAOA circuit.}
    \label{fig:qaoa}
\end{figure}

The quantum approximate optimization algorithm (QAOA) is used to solve combinatorial optimization problems using a hybrid quantum-classical framew\-ork \cite{qaoa}. Many real-world problems can be formulated such that the solutions are \textit{N}-bit binary strings of the form
\begin{equation}
    z = z_1 z_2 \dots z_N,
\end{equation}
which minimize the classical cost function for $m$ clauses,

\begin{equation}
    C(z) = \sum^{m}_{\alpha = 1}C_{\alpha}(z).
\end{equation}

$C_\alpha(z) = 1$ if clause $\alpha$ is satisfied by $z$ and $0$ otherwise \cite{qaoa}. Through the substitution of spin-operators $\sigma^z_i$ for each $z_i$ in $z$, one can build the cost Hamiltonian $H_C$,

\begin{equation}
    H_C = C(\sigma^z_1, \sigma^z_2,\dots,\sigma^z_N).
\end{equation}

The cost hamiltonian for the minimum vertex cost problem is given by:
\begin{equation}
    H_C = A \sum_{(u, v) \in E}(1-x_u)(1-x_v)+B\sum_{v \in V}x_v
\end{equation}
for an appropriate choice of $A$ and $B$ \cite{ham}. These constant terms are introduced because the minimum vertex cover problem contains hard constraints, which are not compatible with the QAOA, which solves only quadratic unconstrained binary optimization problems. The constant terms $A$ and $B$ refer to the weighting constants in the Hamiltonian. $B$ weights the primary objective, minimizing the size of the vertex cover, while $A$ weights a constraint or penalty term, that every edge have at least one of its vertices in the minimum cover. Since QUBO problems are unconstrained by nature, soft constraints in the form of penalty terms, forming soft constraints, are required.

Next, one can define a mixer Hamiltonian $H_M$,

\begin{equation}
    H_M = \sum^N_{j = 1}\sigma^x_j.
\end{equation}
Through the application of layers of alternating cost and mixer Hamiltonians to the initial state $\ket{+}^{\otimes N}$, an equally-weighted superposition of all states in the computational basis, the QAOA circuit from Figure \ref{fig:qaoa} is constructed \cite{qaoa}. This yields

\begin{equation}
    \ket{\boldsymbol{\psi_p}\left(\Vec{\gamma}, \Vec{\beta}\right)} = e^{-i\beta_pH_M}e^{-i\gamma_pH_C} \dots e^{-i\beta_1H_M}e^{-i\gamma_1H_C}\ket{+}^{\otimes N},
\end{equation}
where $p>1$ is the number of layers in the circuit with $2p$ parameters, $\Vec{\gamma_i}$ and $\Vec{\beta_i}$ with $i = 1,2,\dots,p$. A classical optimizer can be used to alter the parameters, to minimize the expectation value,

\begin{equation}
    F_p\left(\Vec{\gamma}, \Vec{\beta}\right) = \bra{\boldsymbol{\psi_p}\left(\Vec{\gamma}, \Vec{\beta}\right)}H_C\ket{\boldsymbol{\psi_p}\left(\Vec{\gamma}, \Vec{\beta}\right)}.
\end{equation}

If $\Vec{\gamma}^* \text{ and } \Vec{\beta}^*$ minimize $F_p$, and if the value of the true solution is given by $z^*$, then the approximation ratio is given by, 
\begin{equation}
    \frac{\left(\Vec{\gamma}, \Vec{\beta}\right)}{z^*}.
\end{equation}

The approximation of the solution $z^*$ can then be obtained through sampling of the state 

\begin{equation}
    \ket{\boldsymbol{\psi_p}\left(\Vec{\gamma^*}, \Vec{\beta^*}\right)},
\end{equation} 
prepared with the optimal parameters $\Vec{\gamma^*}$ and $\Vec{\beta^*}$.

In a fully fault-tolerant setting, the performance of QAOA improves as $p$ increases, however due to limitations in the hardware currently available, the behaviour of QAOA at lower values of $p$ is of great interest \cite{qaoa20}. Some previous works have investigated strategies for improving the performance of QAOA at these small $p$'s, such as using heuristic strategies for selecting the initial parameters, reducing the length of training required \cite{qaoa20}.

\section{Classical Optimizers and Comparisons}
\label{opt}

\subsection{Classical Optimizers}

Variational quantum algorithms, QAOA included, find solution to given problems through the optimization of the ansatz parameters. There are a variety of classical optimizers that can be employed in these variational quantum algorithms. These optimizers can be grouped broadly into two categories: gradient-based and gradient-free. Gradient-based methods use gradient values during the optimization process. The gradient value evaluations can be done analytically for the expectation value of a quantum ansatz on a qubit hamiltonian, through the parameter-shift rule \cite{parameter_shift}, or more primitively, they can be estimated through finite differences. The parameter-shift rule is preferable because the analytic gradient is calculated through much larger variations of the ansatz parameters (and are therefore less susceptible to noise compared to finite differences), while still only requiring the original ansatz circuit in order to calculate the gradient (the same as finite differences). Gradient-free methods require only cost function evaluations, operating as a black-box optimizer. The optimizers compared are listed in Table \ref{tab:my-table}.

\begin{table}[]
\centering
\begin{tabular}{ll}

\textbf{Gradient-Free} & \\                          
Constrained Optimization by Linear Approximation$^{\alpha}$ (COBYLA) & \cite{cobyla} \\
Nelder-Mead$^{\alpha}$ & \cite{nelder-mead}  \\
Modified Powell's method$^{\alpha}$ & \cite{powell}  \\
Simultaneous Perturbation Stochastic Approximation$^{\beta}$ (SPSA)  &\cite{spsa}  \\
\\
\textbf{Gradient-Based} \\
Broyden–Fletcher–Goldfarb–Shanno algorithm$^{\alpha}$ (BFGS) &\cite{bfgs} \\
Limited-memory Broyden–Fletcher–Goldfarb–Shanno algorithm$^{\alpha}$ (L-BFGS) & \cite{l-bfgs}  \\
Conjugate Gradient method$^{\alpha}$ (CG) & \cite{cg} \\
Sequential Least Squares Programming$^{\alpha}$ (SLSQP) & \cite{slsqp} \\
ADAM$^{\beta}$ & \cite{adam} \\
AMSGrad$^{\beta}$ & \cite{amsgrad}\\
\end{tabular}
\caption{Table of Classical Optimizers Considered - The implementation of all classical optimizers are taken from the Scipy$^{(\alpha)}$ and Qiskit$^{(\beta)}$ Python libraries, under the respective functions \textbf{scipy.optimize.minimize} and \textbf{qiskit.aqua.components.optimizers}. (References are given in the far-right column)}
\label{tab:my-table}
\end{table}

This paper utilizes the parameter-shift rule for gradient calculations in all the gradient-based optimizers.

\subsection{Classical Optimizer Comparison}
The comparison of the aforementioned classical optimizers employed in the QAOA on the Min-Vertex Cover problem is described as follows. 

The QAOA algorithm problem test set contains all 21 non-isomorphic 5-vertex graphs. The QAOA is applied to the Min-Vertex Cover problem ten times for each graph in the problem test set. This is repeated for each optimizer and each noise-level. Three levels of noise are considered, all resulting from the type of quantum simulation applied. These simulations are state vector, shot-based fault-tolerant, and shot-based with a sampled noise model, resulting in the noise-levels of noise-free, shot-noise and realistic quantum noise for currently existing quantum computers. The number of shots for the shot-based fault-tolerant and shot-based sampled-noise model is set to 10000. The number of iterations performed by the classical optimizers was limited by the number of function evaluations, which was set to 5000. When utilizing these algorithms on a quantum device, the number of function evaluations has a direct effect on the cost incurred as it determines the amount of device usage. The realistic noise model is sampled from the IBM Belem device. The implementation is done using Pennylane and the Pennylane-Qiskit package, in order to use Qiskit quantum backends and Noise-models.

\subsection{QAOA Ansatz Depth Comparison}

\begin{figure}
    \includegraphics[width=\textwidth]{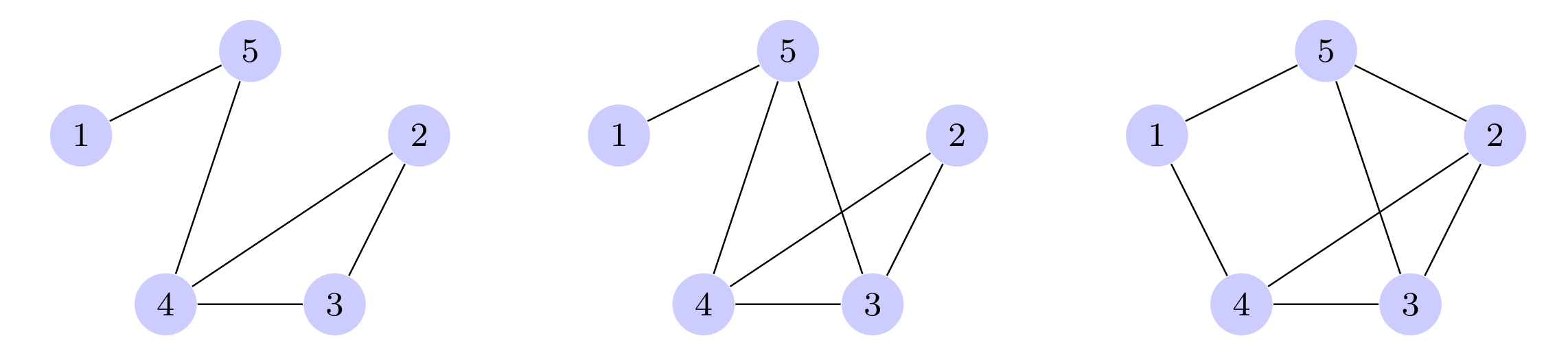}
    \caption{The three graphs used in the QAOA depth comparison, referred to as graph 1, 2 and 3 respectively}
    \label{graphs}
\end{figure}

Once the most suitable classical optimizer is found, the next comparison finds the optimal depth of the QAOA ansatz circuit. In a state vector simulation of the QAOA algorithm, the accuracy increases as the number of layers increases, with the exact answer being achieved at the limit at which the number of layers approaches infinite. On a noisy quantum device, a deeper ansatz circuit will be more affected by noise, and hence the less the simulation will approximate the state vector simulation. This creates a trade-off between the theoretical accuracy increase achieved by increasing the number of layers, with the decrease in accuracy of the simulation caused by the decreased noise resistance that comes with the increased depth of the ansatz. More layers mean a more expressible ansatz (hence a better answer), but also more effects from noise (hence a worse answer). 

For the three graphs in Figure: \ref{graphs}, we run a set of one hundred noisy simulations for each depth ranging from 1 to 10 layers, with the same noise model sampled in the optimizer comparison. We run the same simulations on a state vector simulator to show the theoretical convergence is achieved as the number of layer increases.  

\subsection{QAOA Ansatz Depth Recommendation Verification}

\begin{figure}
    \includegraphics[width=\textwidth]{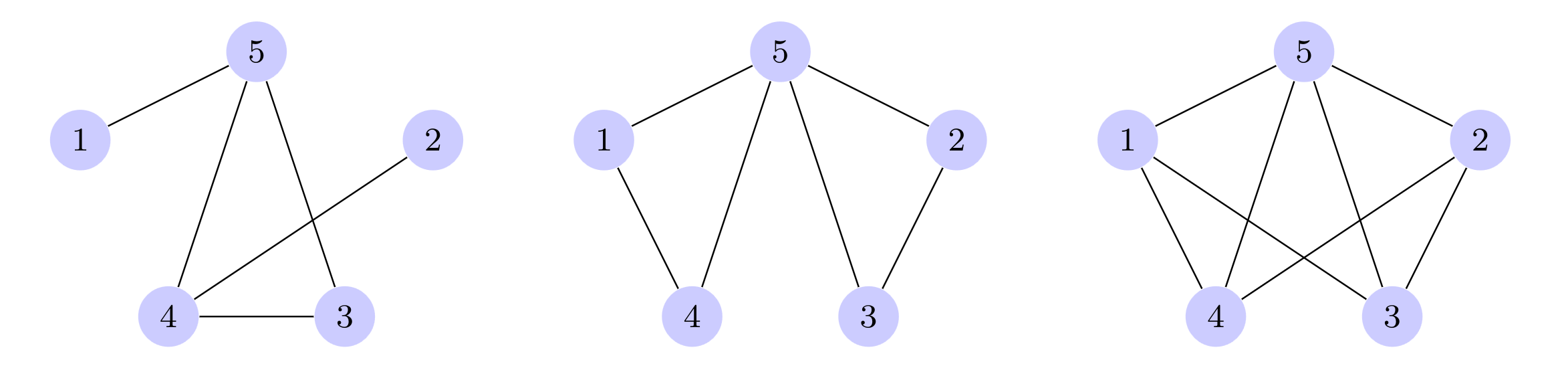}
    \caption{The three graphs used in the QAOA depth recommendation comparison, referred to as graph 4, 5 and 6 respectively}
    \label{graphs 2}
\end{figure}

Once the optimal depth is estimated from the experiments above, we seek to verify that these do indeed maximize the performance of the QAOA Algorithm on the Min-Vertex Cover problem for graphs of this size (5 qubits, 5 edges) in Figure: \ref{graphs 2}. We compare the QAOA with differing numbers of layers, and show that the solutions that are sampled from the QAOA with optimized parameters, are, on average, better when sampled from the QAOA with the optimal number of layers. It is expected that when the optimal number of parameters are used in the QAOA, the solutions sampled from the optimized QAOA ansatz will be on average better. 

\section{Results}
\label{results}
\subsection{Comparison of Classical Optimizers}
\label{part one}

\begin{figure}
    \centering
    \subfloat[][]{
        \includegraphics[width=0.75\textwidth]{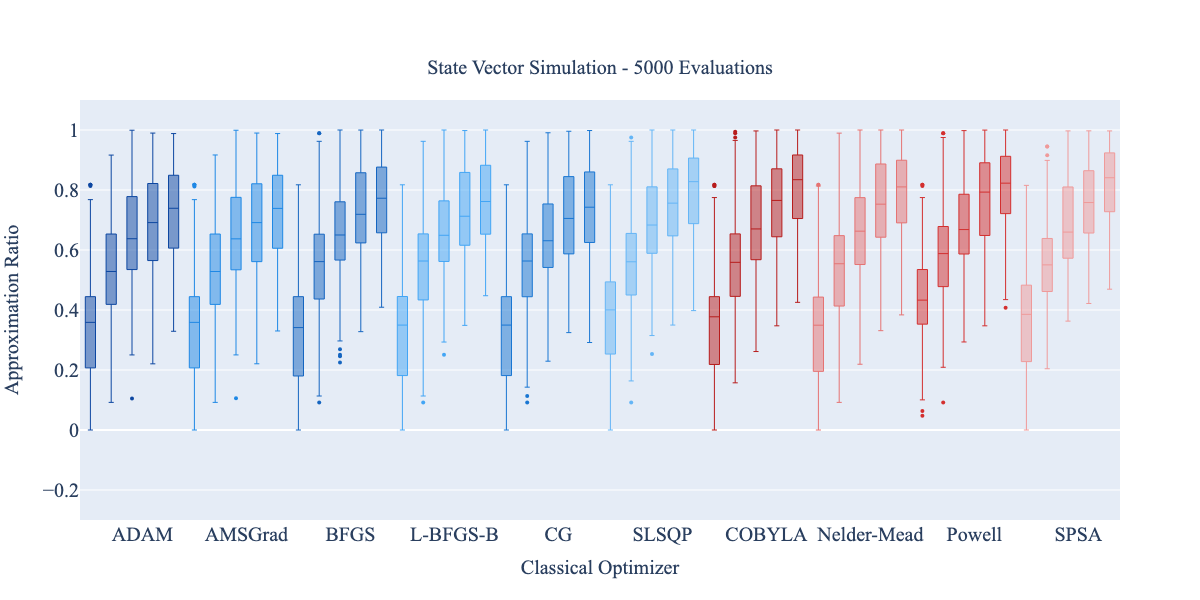}
        \label{sv_comp}
    }
    
    \subfloat[][]{
        \includegraphics[width=0.75\textwidth]{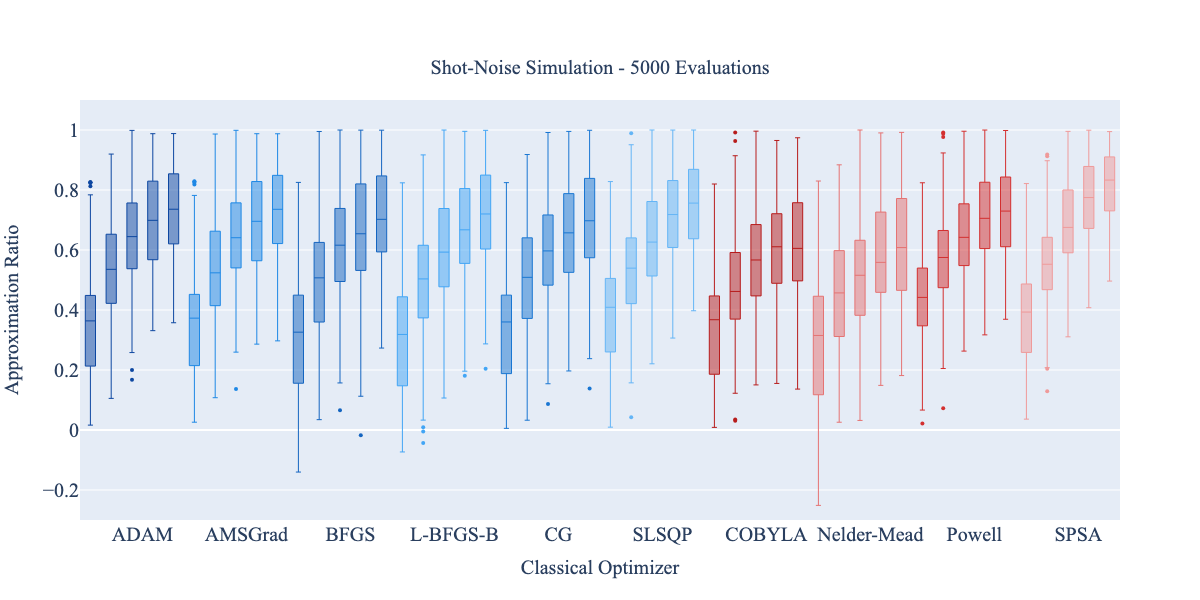}
        \label{shot_comp}
    }
    
    \subfloat[][]{
        \includegraphics[width=0.75\textwidth]{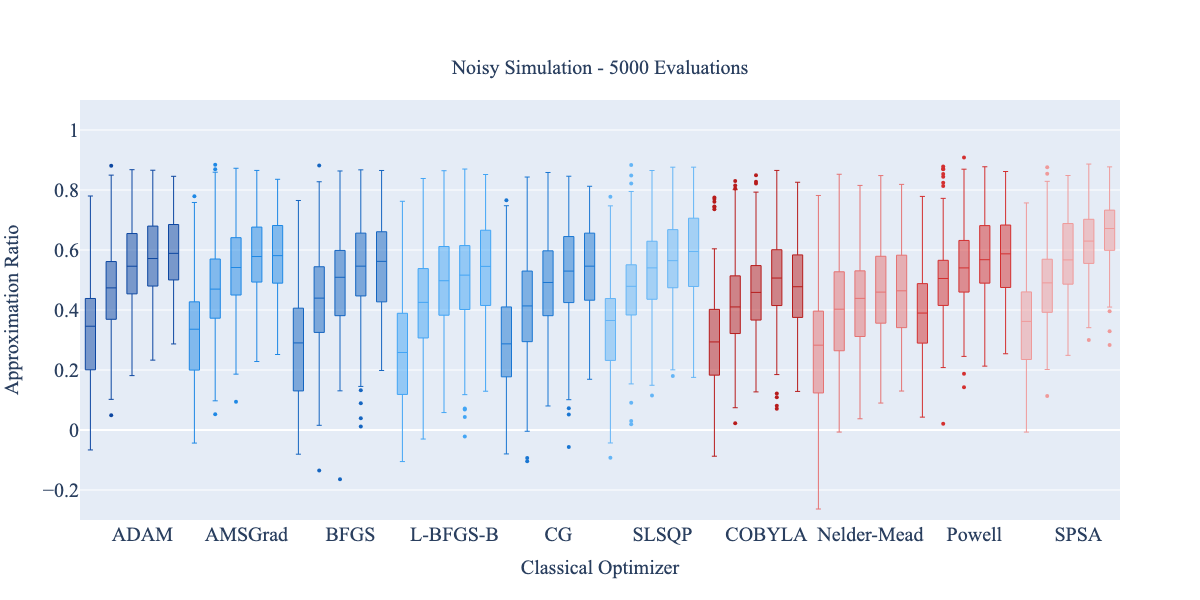}
        \label{noise_comp}
    }
    \caption{A comparison of the approximation ratio using 10 classical optimizers on the QAOA, each having depths 1 to 5 from left to right, using a statevector, shot-based, and noisy simulation in (a), (b) and (c) respectively. Each box is made up of the 10 runs for each of the 21 non-isomorphic graphs. The three graphs capture the change in performance as one moves from application in a noiseless regime, to the application of the methods in the presence of realistic noise from a quantum device. This is demonstrates the change in performance that could be expected on a real device and it can be observed that as more noise is introduced to the system, there is a decrease in the performance of the classical optimizers, with some optimizers being more affected than others. Each problem instance optimization is done with a maximum of 5000 cost function evaluations.}
    \label{optimizer_comparison}
\end{figure}

The results for the different classical optimizers, gradient-free and gradient-based, for each ansatz depth from 1 layer to 5 layers are shown in Figure  \ref{optimizer_comparison} for the state vector, shot-based and noisy simulations in \ref{sv_comp}, \ref{shot_comp} and \ref{noise_comp} respectively. 

In the state vector simulation, all ten optimizers appear to perform similarly, with no major distinction between the gradient-free and gradient-based optimizers. There is a clear trend that more layers yields a better approximation ratio.

When shot-noise is accounted for, the differences between the optimizers become noticeable. SPSA slightly outperforms the other optimizers in this setting. The gradient-based optimizers all perform equally well, equivalent to the performance of Powell. COBYLA and Nelder-Mead are noticeably affected by the inclusion of shot-noise. For COBYLA, the approximation ratio for five layers is equivalent to the approximation ratio achieved with four layers, suggesting that this optimizer had trouble effectively making use of the extra parameters in the ansatz due to the shot-noise. All other optimizers show noticeable improvement with the increase in the number of layers.

Finally, when the realistic noise model is incorporated into the simulation, the differences between classical optimizers become even more pronounced.
SPSA is the best performing optimizer overall, achieving the best 4 and 5 layer approximation ratio in the presence of realistic noise-levels. COBYLA and Nelder-Mead are seriously negatively affected by noise, the approximation ratio for 5 layers is worse than that with 4 layers. All gradient-based methods show similar performance, equivalent to that of Powell.

Following these results, SPSA is used to run the noisy simulations for the QAOA ansatz depth comparison in Section \ref{part two} and the ansatz depth verification in Section \ref{part three}. COBYLA is used for the accompanying state vector simulations as it  was found to be the best optimizer in the state vector simulation.

As an interesting note, when limiting the number of function evaluations permitted by the classical optimizer, the gradient-based optimizers' ability to accurately converge is severely affected. Thus, when utilizing algorithms with an ansatz based on Hamiltonian simulation on quantum hardware, it is not recommended that one make use of these gradient-based optimizers. As these optimizers require more calls to the quantum device to evaluate the gradient, limiting this causes the observed degradation in performance. For the different types of variational circuits similar results were reported in \cite{Grimsley,Solinas}. 

\subsection{QAOA Ansatz Depth Comparison}
\label{part two}

\begin{figure}
    \centering
    \subfloat[][]{
        \includegraphics[width=0.8\textwidth]{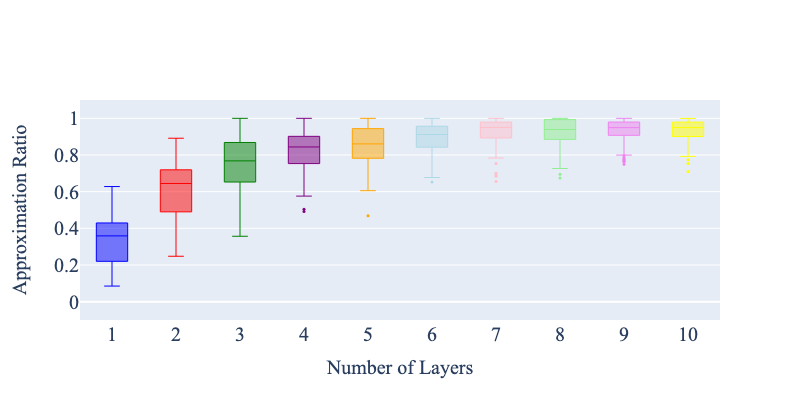}
    }
    
    \subfloat[][]{
        \includegraphics[width=0.8\textwidth]{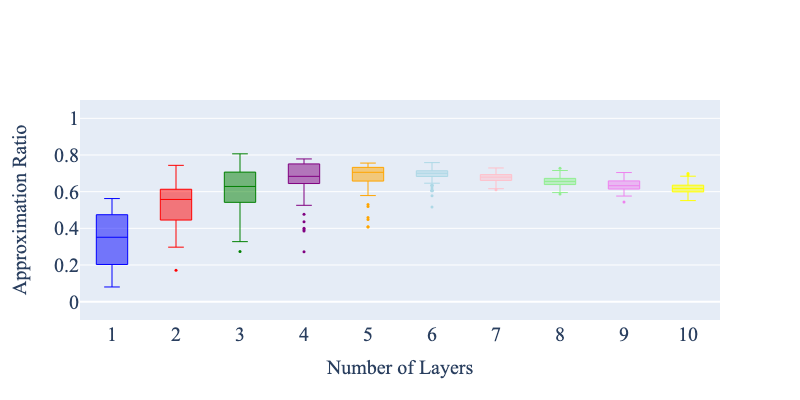}
    }
    
    \caption{Number of layers vs approximation ratio of a state vector (a) and noisy (b) simulation of QAOA for the minimum vertex cover on graph 1, with a maximum of 5000 cost function evaluations per problem instance. Each bar represents 100 runs of the QAOA for each number of layers, for the same minimum vertex cover problem.}
    \label{depth graph 0}
\end{figure}

\begin{figure}
    \centering
    \subfloat[][]{
        \includegraphics[width=0.8\textwidth]{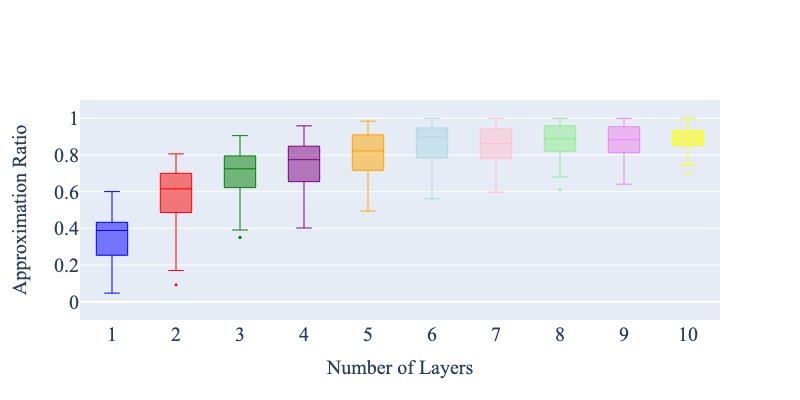}
    }
    
    \subfloat[][]{
        \includegraphics[width=0.8\textwidth]{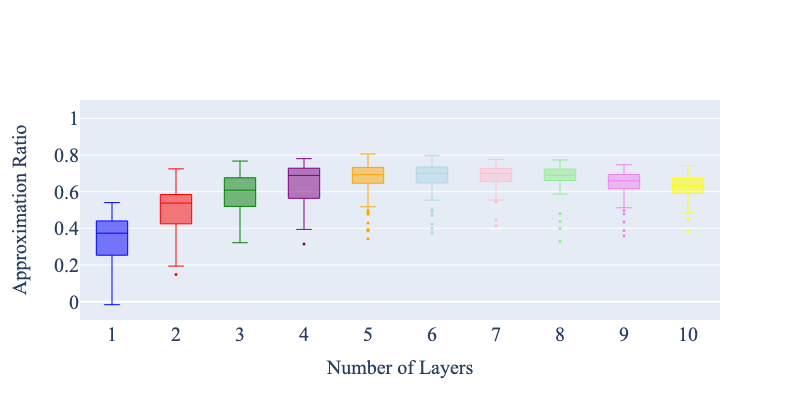}
    }
    
    \caption{Number of layers vs approximation ratio of a state vector (a) and noisy (b) simulation of QAOA for the minimum vertex cover on graph 2, with a maximum of 5000 cost function evaluations per problem instance. Each bar represents 100 runs of the QAOA for each number of layers, for the same minimum vertex cover problem.}
    \label{depth graph 1}
\end{figure}

\begin{figure}
    \centering
    \subfloat[][]{
        \includegraphics[width=0.8\textwidth]{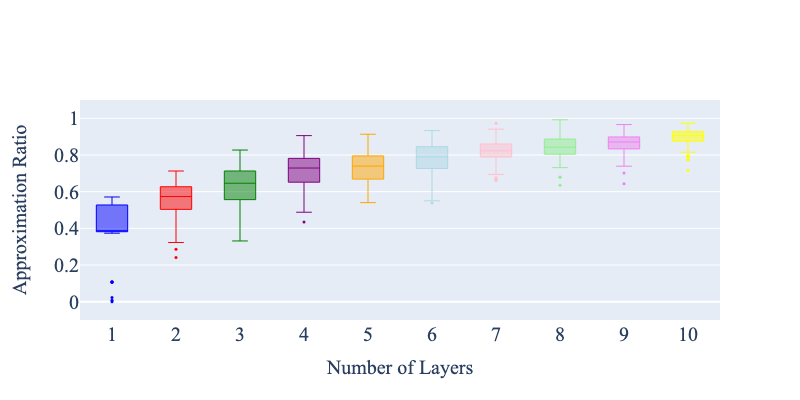}
    }
    
    \subfloat[][]{
        \includegraphics[width=0.8\textwidth]{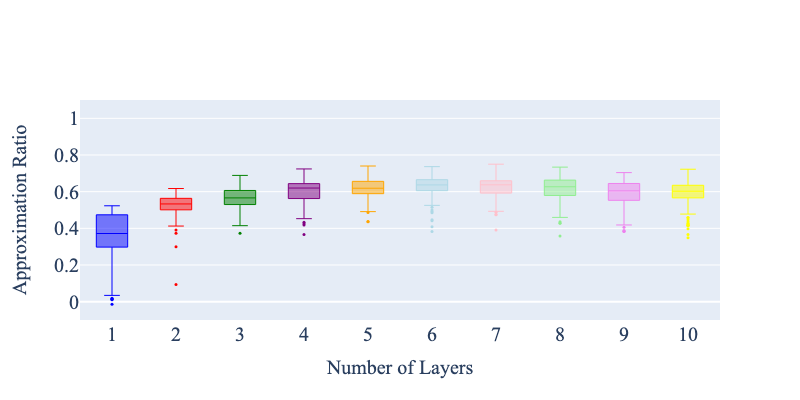}
    }
    
    \caption{Number of layers vs approximation ratio of a state vector (a) and noisy (b) simulation of QAOA for the minimum vertex cover on graph 3, with a maximum of 5000 cost function evaluations per problem instance. Each bar represents 100 runs of the QAOA for each number of layers, for the same minimum vertex cover problem.}
    \label{depth graph 2}
\end{figure}

Figures \ref{depth graph 0}, \ref{depth graph 1} and \ref{depth graph 2}, give the comparison between the state vector and noisy simulations for the three graphs respectively. 

In the state vector simulations, the expected trend is clearly apparent; that the approximation ratio of the QAOA algorithm improves steadily as additional layers are added. It is also clear that adding another layer seems to improve the approximation ratio less than the addition of the layer before. There is a diminishing return on accuracy increase with the number of layers added.

In the noisy simulations, as the number of layers increases, the effect of noise in the simulation becomes apparent. With the best approximation ratio achieved at around six layers for all graphs. Once this best approximation ratio is achieved, additional layers then begin to slowly worsen the approximation ratio. 

\subsection{QAOA Ansatz Depth Recommendation Verification}
\label{part three}

\begin{figure}
    \centering
    \subfloat[][]{
        \includegraphics[width=0.95\textwidth]{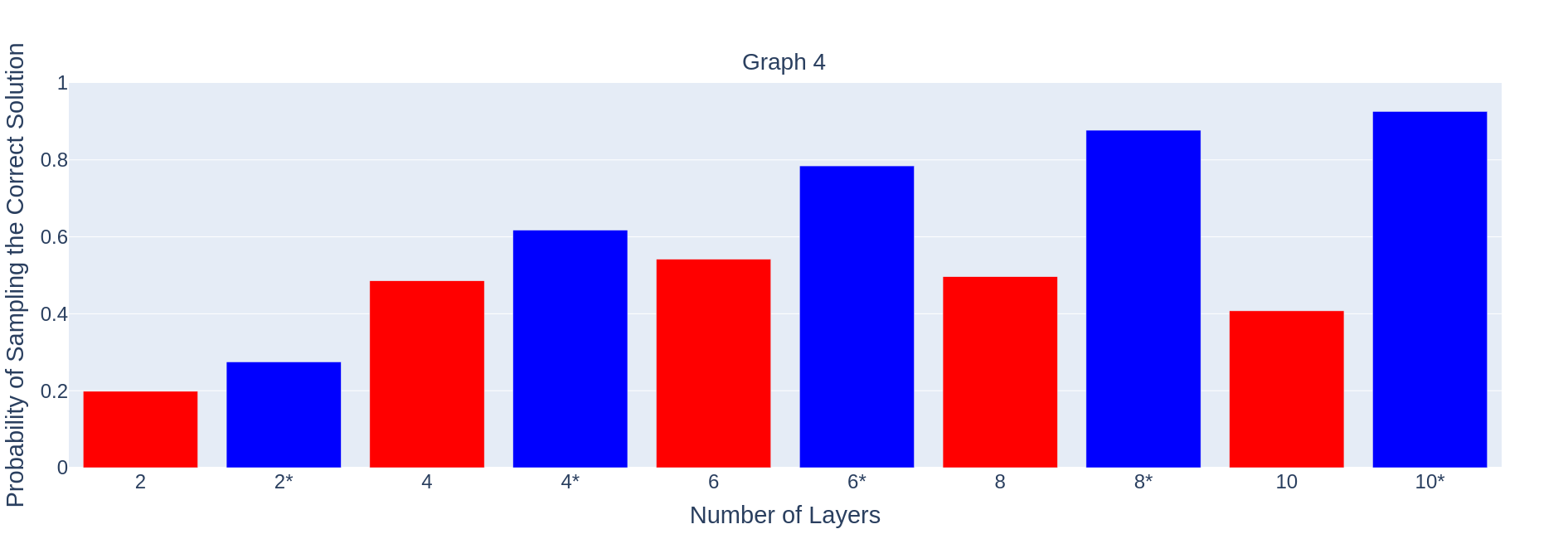}
        \label{fig:graph_4}
    }
    
    \subfloat[][]{
        \includegraphics[width=0.95\textwidth]{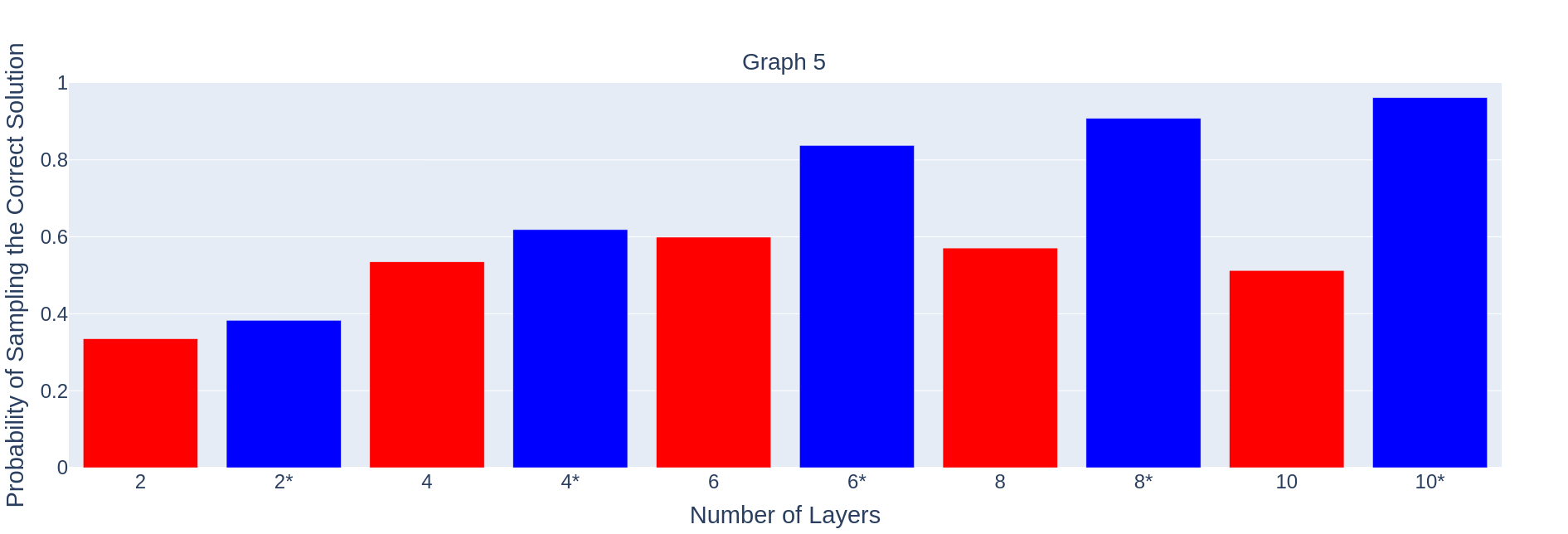}
        \label{fig:graph_5}
    }

    \subfloat[][]{
        \includegraphics[width=0.95\textwidth]{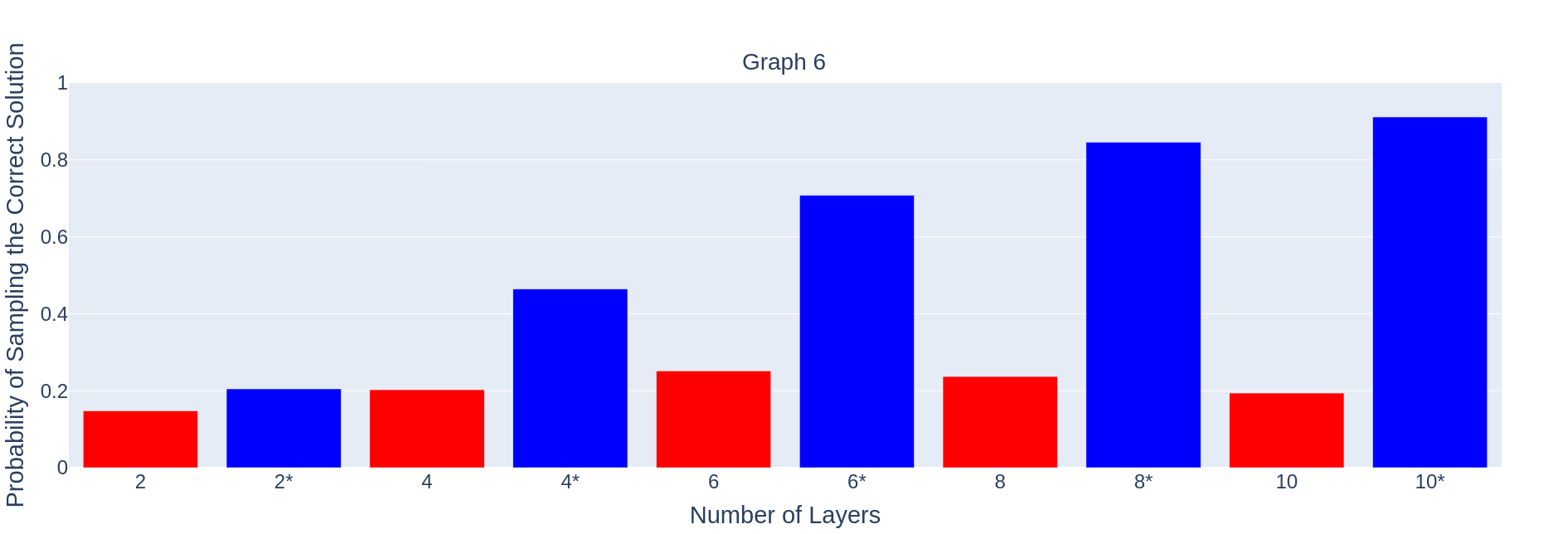}
        \label{fig:graph_6}
    }
    
    \caption{Number of layers vs probability of sampling the correct solution to the minimum vertex cover problem on each graph respectively (a, b, c) , for both noisy (red) and statevector (blue) simulations.}
    \label{depth graph end}
\end{figure}

Figures \ref{fig:graph_4}, \ref{fig:graph_5}, \ref{fig:graph_6} show the probabilities of sampling the correct solution for the minimum vertex cover problem on graphs 4, 5 and 6, for the noise and state vector simulation respectively.

It is clear from these graphs, that 6 layers appears to be optimal for these instances of the QAOA too, allowing the correct solution for the minimum vertex cover on these graphs to be sampled with the greatest probability. After 6 layers, additional layers decrease the probability of sampling the correct solution for the minimum vertex cover problem.

\section{Discussion and Conclusion}

The results from Section \ref{part one}, the comparison of classical optimizers in the QAOA, show that the choice of classical optimizer has a significant effect on the algorithm in the presence of noise. A classical optimizer's performance in a state vector simulation does not accurately reflect its performance in a realistic noise setting. It appears that SPSA is the best classical optimizer for current levels of noise. This is a result of both its built-in stochastic nature making it more resistant to noise, and its efficient gradient approximation requiring only two cost function evaluations for any ansatz. These results are similar to those found in \cite{pellow-jarman}, where SPSA was found to be the best classical optimizer in the noisy simulation, while COBYLA and Nelder-Mead were found to be the worst.

The results from Section \ref{part two}, the comparison of depths for the QAOA ansatz, show that while in theory the more layers present in the QAOA the greater the accuracy, it is actually the case that in the presence of noise in the circuit, there is actually an optimal number of layers that provide the greatest accuracy. It is therefore important to use the correct number of layers in order to utilize the QAOA algorithm to its full potential. Previous works have suggested other guidelines for ansatz depth based on factors such as time complexity of execution \cite{hadfield}. Section \ref{part three} shows that the probability of sampling the correct solution for the QAOA problem is also greatest at the optimal number of layers, and decrease as the number of layers increase beyond that.

It is left as a future work to fully characterize the trade-off level of noise in the circuit and the accuracy improvement yielded by adding additional layers to the QAOA. It is hoped that it will then be possible to estimate the optimal number of layers for a QAOA circuit for a given problem and a device's noise level.

\section*{Acknowledgments}
\label{ACKNOWLEDGMENTS}
Support from the NICIS (National Integrated Cyber Infrastructure System) e-research grant QICSI7 is kindly acknowledged. This research is also supported by the Yonsei University Research Fund of 2024 (2024-22-0147), the National Research Foundation of Korea (Grant No. 2022M3E4A1074591), and the KIST Institutional Program (2E32941-24-008).

\section*{Data Availability}
The data is available at \textbf{https://github.com/aidanpellow/qaoa}.

\section*{Author contributions statement}

D.K.P., A.P and I.S. conceived the experiments,  A.P. conducted the experiment(s), A.P. and I.S. analysed the results. A.P., S.M. and I.S. wrote the paper.  All authors reviewed the manuscript. 

\section*{Additional information}

\textbf{Competing interests} Francesco Petruccione the Chair of the Scientific Board and Co-Founder of QUNOVA computing. Aidan Pellow-Jarman and Shane McFarthing are employees of Qunova Computing.  The authors declare no other competing interests.

\begin{thebibliography}{}

\bibitem{nielsen}
Nielsen, M. A., and Chuang I., ``Quantum computation and quantum information," (2002).

\bibitem{grover}
Grover, L. K., ``A fast quantum mechanical algorithm for database search," In Proceedings of the twenty-eighth annual ACM symposium on Theory of computing, Pp. 212-219 (1996).

\bibitem{shor}
Shor, Peter W., ``Algorithms for quantum computation: discrete logarithms and factoring," In Proceedings 35th annual symposium on foundations of computer science, Ieee, Pp. 124-134 (1994).

\bibitem{preskill}
Preskill, J., ``Quantum computing in the NISQ era and beyond," Quantum, Iss. 2, Pp. 79 (2018).

\bibitem{peruzzo}
Peruzzo A, et al. ``A variational eigenvalue solver on a photonic quantum processor," Nature communications, Vol. 5, Iss. 1, Pp. 4213 (2014).

\bibitem{qaoa}
E. Farhi and J. Goldstone, ``A Quantum Approximate Optimization Algorithm," (2014) \textbf{arXiv:1411.4028}

\bibitem{qaoa_opt}
Fern\`{a}ndez-Pend\`{a}s M, et. al., ``A study of the performance of classical minimizers in the Quantum Approximate Optimization Algorithm," Journal of Computational and Applied Mathematics, 404 (2022).

\bibitem{pellow-jarman}
Pellow-Jarman, A, et al., ``A comparison of various classical optimizers for a variational quantum linear solver," Quantum Information Processing, Vol. 20, Iss. 6, Pp. 202 (2021).

\bibitem{qaoa_var}
Wenyang Qian et.al., ``Comparitive study of variations in quantum approximate optimization algorithms for the Traveling Salesman Problem \textbf{arXiv:2307.07243}"

\bibitem{qaoa20}
Zhou, L., Wang, S. T., Choi, S., Pichler, H., and Lukin, M. D., ``Quantum approximate optimization algorithm: Performance, mechanism, and implementation on near-term devices," Physical Review X, Vol. 10, Iss. 2, Pp. 021067 (2020).

\bibitem{ham}
Pelofske, E., Hahn, G. and Djidjev, H., ``Decomposition Algorithms for Solving NP-hard Problems on a Quantum Annealer", Journal of Signal Processing Systems, Vol. 93, Pp 405-420 (2021)

\bibitem{parameter_shift}
Wierichs D., Izaac J., Wang C., and Yen-Yu Lin C., ``General parameter-shift rules for quantum gradients," Quantum 6, 677 (2022)

\bibitem{cobyla}
M. J. D. Powell, ``A direct search optimization method that models the objective and constraint functions by linear interpolation," Advances in Optimization and Numerical Analysis, Pp. 51-67 (1994)

\bibitem{nelder-mead}
J. A. Nelder and R. Mead, ``A simplex method for function minimization," The Computer Journal, Vol. 7, Iss. 4, Pp. 308–313 (1965)

\bibitem{powell}
M. J. D. Powell, ``An efficient method for finding the minimum of a function of several variables without calculating derivatives," Computer Journal, Vol. 7, Iss. 2, Pp. 155–162 (1964)

\bibitem{spsa}
J. C. Spall. ``Multivariate Stochastic Approximation Using a Simultaneous Perturbation Gradient Approximation," IEEE Transactions on Automatic Control, Vol. 37, No. 3, Pp. 332–341, (1992)

\bibitem{bfgs}
R. Fletcher, ``A  new  approach  to  variable  metric  algorithms," The Computer Journal, Vol. 13, Iss. 3, Pp 317–322 (1970)

\bibitem{l-bfgs}
Richard H. Byrd, Peihuang Lu, Jorge Nocedal and Ciyou Zhu, ``A Limited Memory Algorithm for Bound Constrained Optimization," SIAM Journal on Scientific Computing, Vol. 16, Pp. 1190-1208 (1995)

\bibitem{cg}
Magnus R. Hestenes and Eduard Stiefel, ``Methods of Conjugate Gradients for Solving Linear Systems," Journal of Research of the National Bureau of Standards, Vol. 49, No. 6 (1952)

\bibitem{slsqp}
Paul T. Boggs, Jon W. Tolle, ``Sequential Quadratic Programming," Acta Numerica, Vol. 4, Pp 1-51 (1996)  

\bibitem{adam}
Diederik P. Kingma and Jimmy Lei Ba, ``ADAM: A Method for Stochastic Optimization" (2017) \textbf{arXiv:1412.6980}
    
\bibitem{amsgrad}
Sashank J. Reddi, Satyen Kale and Sanjiv Kumar, ``On the Convergence of ADAM and Beyond" (2018) \textbf{arXiv:1904.09237}

\bibitem{Grimsley}
Grimsley, H.R., Economou, S.E., Barnes, E. et al. ``An adaptive variational algorithm for exact molecular simulations on a quantum computer", Nat Commun 10, 3007 (2019). 

\bibitem{Solinas}
Solinas, P., Caletti, S. and Minuto, G. ``Quantum gradient evaluation through quantum non-demolition measurements", Eur. Phys. J. D 77, 76 (2023). 

\bibitem{hadfield}
Hadfield, Stuart, Zhihui Wang, Bryan O’Gorman, Eleanor G. Rieffel, Davide Venturelli, and Rupak Biswas, ``From the Quantum Approximate Optimization Algorithm to a Quantum Alternating Operator Ansatz", Algorithms, Vol 12(2):34, Pp 1- 45 (2019)

\end{thebibliography}
\end{document}